\documentclass[12pt]{article}

\usepackage{scicite}
\usepackage[pdftex]{graphicx}

\newenvironment{sciabstract}{%
\begin{quote} \bf}
{\end{quote}}

\newcounter{lastnote}

\title{Pseudo Three-dimensional Maps of the Diffuse Interstellar Band at 862 nm} 

\author{
Janez Kos,$^{1\ast}$ \and Toma\v{z} Zwitter,$^{1}$ \and Rosemary Wyse$^{2}$ \and Olivier Bienaym\'{e},$^{3}$ \and
James Binney,$^{4}$ \and Joss Bland-Hawthorn,$^{5}$ \and Kenneth Freeman,$^{6}$ \and Brad K. Gibson,$^{7}$ \and Gerry Gilmore,$^{8}$ \and Eva K. Grebel,$^{9}$ \and Amina Helmi,$^{10}$ \and Georges Kordopatis,$^{8}$ \and Ulisse Munari,$^{11}$ \and Julio Navarro,$^{12}$ \and Quentin Parker,$^{13, 14, 15}$ \and Warren A. Reid,$^{13, 14}$ \and George Seabroke,$^{16}$ \and Sanjib Sharma,$^{5}$ \and Arnaud Siebert,$^{3}$ \and Alessandro Siviero,$^{17, 18}$ \and Matthias Steinmetz,$^{18}$ \and Fred G. Watson,$^{15}$ \and Mary E. K. Williams,$^{18}$}

\date{}

\begin{document} 


\baselineskip24pt


\maketitle 

\begin{center}
\normalsize{$^1$Faculty of Mathematics and Physics, University of Ljubljana, Jadranska 19, 1000 Ljubljana, Slovenia,}\\
\normalsize{$^{2}$Johns Hopkins University, Homewood Campus, 3400 N Charles Street, Baltimore, MD 21218, USA,}\\
\normalsize{$^3$Observatoire astronomique de Strasbourg, Universit\'e de Strasbourg, CNRS, 11 rue de l'Universit\'e, F-67000 Strasbourg, France,}\\
\normalsize{$^4$Rudolf Peierls Centre for Theoretical Physics, Keble Road, Oxford OX1 3NP, UK,}\\
\normalsize{$^5$Sydney Institute for Astronomy, School of Physics A28, University of Sydney, NSW 2008, Australia,}\\
\normalsize{$^6$Research School of Astronomy \& Astrophysics, Australian National University, Canberra, Australia,}\\
\normalsize{$^7$Chair, Computational Astrophysics, Jeremiah Horrocks Institute, University of Central Lancashire, Preston, PR1 2HE, United Kingdom,}\\
\normalsize{$^8$Institute of Astronomy, Madingley Road, Cambridge CB3 0HA, UK,}\\
\normalsize{$^9$Astronomisches Rechen-Institut, Zentrum f\"ur Astronomie der Universit\"at Heidelberg, M\"onchhofstra\ss e 12 -- 14, D-69120 Heidelberg, Germany,}\\
\normalsize{$^{10}$Kapteyn Astronomical Institute, PO Box 800, NL-9700 AV Groningen, the Netherlands,}\\
\normalsize{$^{11}$INAF Astronomical Observatory of Padova, 36012 Asiago (VI), Italy,}\\
\normalsize{$^{12}$Senior ClfAR Fellow. University of Victoria, Victoria BC, Canada V8P 5C2,}\\
\normalsize{$^{13}$Department of Physics and Astronomy, Macquarie University, Sydney, NSW 2109, Australia,}\\
\normalsize{$^{14}$Centre for Astronomy, Astrophysics and Astrophotonics, Macquarie University, Sydney, NSW 2109, Australia,}\\
\normalsize{$^{15}$Australian Astronomical Observatory, PO Box 915, North Ryde, NSW 1670,}\\
\normalsize{$^{16}$Mullard Space Science Laboratory, University College London, Holmbury St Mary, Dorking, RH5 6NT, UK,}\\
\normalsize{$^{17}$Department of Physics and Astronomy, Padova University, Vicolo dell'Osservatorio 2, I-35122 Padova, Italy,}\\
\normalsize{$^{18}$Leibniz-Institut f\"ur Astrophysik Potsdam (AIP), An der Sternwarte 16, 14482 Potsdam, Germany,}\\
\normalsize{$^\ast$To whom correspondence should be addressed; E-mail:  janez.kos@fmf.uni-lj.si.}
\end{center}


\begin{sciabstract}
The diffuse interstellar bands (DIBs) are absorption lines observed in visual and near infrared spectra of stars. Understanding their origin in the interstellar medium (ISM) is one of the oldest problems in astronomical spectroscopy, as DIBs ahave been known since 1922. In a completely new approach to understanding DIBs, we combined information from nearly 500,000 stellar spectra obtained by the massive spectroscopic survey RAVE (Radial Velocity Experiment) to produce the first pseudo three-dimensional map of the strength of the DIB at 8620~{\AA}ngstroms covering the nearest 3 kiloparsecs from the Sun, and show that it follows our independently constructed spatial distribution of extinction by interstellar dust along the Galactic plane. Despite having a similar distribution in the Galactic plane, the DIB 8620 carrier has a significantly larger vertical scale height than the dust. Even if one DIB may not represent the general DIB population, our observations outline the future direction of DIB research.
\end{sciabstract}

Diffuse Instellar Bands (DIBs) are wide and sometimes structured absorption lines in the optical and near-infrared wavelengths that originate in the interstellar medium (ISM) and were discovered in 1922\cite{herbig95,sarre06};  more than 400 are known today\cite{hobbs09}, but their physical carriers are still unidentified\cite{gala11, krelowski10, salama99, iglesias10,maier11}. Their abundances are correlated with interstellar extinction and with abundances of some simple molecules\cite{thorburn03}, so DIBs are probably associated with carbon-based molecules\cite{sarre06}. DIBs show no polarization effects\cite{herbig95} and are likely positively charged\cite{mili14}, as suggested by the relatively low energies of absorbed photons\cite{tielens}. No known transition of any molecule or atom has yet been found to match the central wavelengths of the DIBs\cite{sarre06}. Their origin and chemistry are thus unknown, a unique situation given the distinctive family of many absorption lines within a limited spectral range. Like most molecules in the ISM that have an interlaced chemistry, DIBs may play an important role in the life-cycle of the ISM species and are the last step to fully understanding the basic components of the ISM. The problem of their identity is more intriguing given the possibility that the DIB carriers are organic molecules. DIBs remain a puzzle for astronomers studying the ISM, physicists interested in molecular spectra, and chemists studying possible carriers in the laboratories.

The maps presented here are based on data from the recently completed RAV (Radial velocity experiment)E spectroscopic survey\cite{kordopatis13} and are limited to one DIB. However extensive surveys of Galactic stars \cite{eisenstein11, gilmore12, gaia12, galah12, deng12} that are starting now will permit spatial mapping, the study of environmental constraints, and a comparison of the spatial distribution for over a dozen DIBs, interstellar molecules, and dust with techniques similar to the ones described here. Large spectroscopic surveys observing $\sim10^5$ of stars are a big leap forward from previous DIB-specific surveys, where only a few thousand stars were observed at best\cite{snow77,loon13}, or only around a hundred stars, if weaker DIBs were observed\cite{friedman11}. 

Making a three--dimensional (3D) map of an ISM species from absorption lines in stellar spectra is a challenge, requiring a large number of observed lines of sight with measured abundances of the ISM species as well as the distances to the observed stars. The DIB carrier can be found anywhere along the line of sight to the observed star, so the distance to the DIB carrier is uncertain, with only an upper limit from the distance to the star. It can only be established through the observations of many stars within a small solid angle but at different distances. To produce a map, such sets of observations must be achievable in  any direction. The RAVE survey fulfills these requirements. In the pre-Gaia era, we have precise distances only to Hipparcos stars and a few others. For maps to be made at distances larger than the Hipparcos sample, spatial resolution must be sacrificed, as the distance measurements are less precise and errors in distance calculations will exceed the typical size of the ISM clumps.

DIBs are more numerous than absorption lines of other ISM species in the optical and NIR bands and are therefore ideal to be studied in general spectroscopic surveys, as they are present in almost any band observed by the surveys mentioned above. Having observations of multiple DIBs also allows the study of different parameters\cite{kos13} of the ISM apart from observing the spatial distribution of a single species. 

DIBs are traditionally observed in the spectra of hot stars, where interstellar lines rarely blend with stellar ones, but hot stars are intrinsically rare. With new analytical methods\cite{kos13b,chen12}, it is possible to observe DIBs in the spectra of cool stars, which dominate in most spectroscopic surveys. In the (magnitude-limited) RAVE survey, only 1.1 \% of the stars have effective temperature above 8000~K. Our method makes use of a large database of RAVE spectra and requires neither knowledge of stellar parameters nor the use of synthetic spectra\cite{kos13b}. For each spectrum, a number of most similar stellar spectra were found in regions with very low extinction at high Galactic latitudes. From these spectra, we generated a stellar template to divide out the stellar contribution in the target spectrum. This leaves only the interstellar features in the spectrum: in this case only the DIB at 862~nm. See \cite{kos13b} for a detailed description of the DIB extraction method. \cite{kos13b} also shows that the RAVE data satisfiy the prerequisities for the analysis of this paper: The DIB~8620 can be detected  after the combination of several RAVE spectra; it can be detected at high Galactic latitudes; and the DIB strength correlates with the interstellar extinction. The DIB in \cite{kos13b} is measured more precisely than the extinction, as is reflected in the correlation plots.

Distances and extinctions are calculated jointly by a Bayesian algorithm \cite{binney13}. As input it takes photometric data and spectroscopic parameters of stellar atmospheres measured in RAVE survey, and it assumes asymptotic values of extinction from the SFD maps\cite{schlegel98}, corrected accordingly to known deviations. The value of the total V-band extinction, $A_V$, is the calculated parameter. (See \cite{sm} for a more detailed description.)

The number of observed stars in the RAVE proved to be high enough for this study only in the nearest 3~kpc of the Galaxy \cite{sm}. The signal-to-noise (S/N) ratio of an individual RAVE spectrum is too low (the mode of the S/N values is 25) to detect the DIB in the spectrum of an individual star, so several spectra were combined to achieve a S/N ratio of $\sim$300. The  DIB is generally detectable in these combined spectra and its strength can be measured with a precision of 10 to 20\% (for details on combining spectra see \cite{kos13b}). This requirement to combine spectra is the limiting factor in the achievable spatial resolution. The resolution of a true 3D map would be low, so we produced a pseudo-3D map, where the distribution of the carrier in the $z$ direction (perpendicular to the Galactic plane) is described by an exponential law with a fixed scale height and a variable in-plane scaling factor, represented by a 2D map separately for the northern and the southern Galactic hemisphere (Fig. 1). Because of the highly variable star density in different volumes, the final maps vary in spatial resolution, between 75~pc and 400~pc. This allowed us to cover a wider volume of the Galaxy than we could with a better, but fixed, resolution. Combining spectra in a given bin also improved the distance value of each bin:  Distance errors of 25\% for individual stars are reduced when averaging over all stars in a bin, so that the error on the bin distance becomes smaller than the bin size.

An exponential law is a good approximation of the spatial distribution of the interstellar dust, as well as the strength of the DIB at 862~nm in the direction perpendicular to the Galactic plane (Fig. 2). The scale heights of the DIB and dust components differ significantly: 117.7$\pm$4.7~pc for the dust and 209.0$\pm$11.9~pc for the DIB bearing gas. These two scale heights were used as constants uniformly through the whole region of the Galaxy included in this study. Some of the gas components, such as H$_2$ or CO (87~pc)\cite{sanders84,wouterloot90} and probably Na~I ($<$200~pc)\cite{sembach93} are consistent with the vertical distribution of the dust layer that we find, while a larger thickness of the DIB layer is similar to the one of Ca~II ($>$200~pc)\cite{sembach93}. H~I has a profile of many components with an average HWHM (half width at half maxium) of 115~pc and large fluctuations\cite{dickey90}. Massive stars, with a high ultraviolet luminosity, have a scale height even smaller than dust\cite{sharma11}.

The strength of the DIB is measured as its equivalent width. The projected equivalent width has been normalized by an exponential law \cite{sm} and represents the equivalent width that would be measured if both the observer and the star were in the Galactic plane. The resulting maps show some distinctive features (Fig. 3). Most notable are large narrow cavities, where the column density is low. In-between these under-dense regions lie clouds of the ISM, some with gentle inceases in  column density and some with well defined boundaries. The smoother transitions are probably due to several less prominent clouds at different Galactic latitudes that collectively produce a gradient in that direction. We could recognize some known absorbing clouds \cite{sm}, but the identification of other features is more difficult, as all the information in the $z$ direction is condensed into two data-points, one for each hemisphere. Although spiral arms are elusive, a steeper rise in density is observed toward the longitudes between $320^\circ$ and $60^\circ$, toward the Sagittarius arm and the Orion spur. We note that the projected equivalent width can decrease with distance. Different bins can represent stars at different Galactic latitudes and the stars included in one bin can have more ISM in front of them than stars included in a more distant bin at the same Galactic longitude. However, the general rise that we find in the DIB equivalent width with distance increases confidence in the maps. 

Because this work only studies one DIB, the detailed results should not be generalized to other DIBs. It is known, from other studies, that different DIBs show different behaviour. The main difference that is expected among other DIBs is in the value of the vertical scale-height and not so much in the projected equivalent width, because the former has little influence on the quite good correlation between DIBs and the interstellar extinction.

The projected distribution of the DIB-bearing gas (Fig. 3) is the first plot of its kind, as it is the only map of any DIB carrier at this scale and the only one taking the distance information as a major parameter. Together with the measured scale height, this is the first 3D study of the spatial distribution of the DIB-bearing ISM clouds. The projected distribution of the extinction due to the interstellar dust is markedly similar to that of the DIB carrier (see \cite{sm} for the correlation analysis), confirming the strong correspondence between the two\cite{munari08}. The map of the extinction is itself an advance, as it maps the regions out of the Galactic plane and probes dust to greater distances than present maps\cite{gontcharov10} of these regions and is consistent with maps in the literature. Our success in producing the maps of the DIB carrier implies  good prospects for future spectroscopic surveys\cite{gilmore12, galah12, gaia12} that will produce similar\cite{gaia12} or better quality\cite{gilmore12, galah12} spectra and will also rely on DIBs to provide information about the ISM. Our work opens new possibilities in the study of DIBs and also offers a unique way of comparing DIBs with other interstellar species by studying their out-of-plane distribution. This can be translated into the study of physical and chemical properties of DIB carriers in the near future.

The measured 3D distribution, especially the unexpectedly high scale-height of the DIB~8620 carrier calls, for a theoretical explanation. There are two options, either the DIB carriers migrate to their observed distances from the Galactic plane, or they are created at these large distances,  from components of the ISM having a similar distribution. The latter is simpler to discuss, as it does not require  knowledge of the chemistry of the DIB carrier or processes in which the carriers are involved. \cite{khoperskov14} showed that mechanisms responsible for dust migration to high altitudes above the Galactic plane segregate small dust particles from large ones, so the small ones form a thicker disk. This is also consistent with the observations of the extinction and reddening at high Galactic latitudes \cite{peek13}.

\bibliographystyle{science}
\bibliography{dibbib}

\section*{Acknowledgements}
The authors are grateful to the anonymous referees for their valuable advice on the discussion of possible DIB origins. JK and TZ wish to thank A. F. Kodre for a stimulating discussion of the topic.
Funding for RAVE has been provided by: the Anglo-Australian Observatory;
the Leibniz-Institut f\"ur Astrophysik Potsdam; the Australian National
University; the Australian Research Council; the French National Research
Agency; the German Research Foundation (SFB 881); the Istituto Nazionale di
Astrofisica at Padova; The Johns Hopkins University; the National Science
Foundation of the USA (AST-0908326); the W.M. Keck foundation; the
Macquarie University; the Netherlands Research School for Astronomy; the
Natural Sciences and Engineering Research Council of Canada; the Slovenian
Research Agency; Center of Excellence Space.si; the Swiss National Science
Foundation; the Science \& Technology Facilities Council of the UK;
Opticon; Strasbourg Observatory; European Research Council; and the Universities of Groningen,
Heidelberg and Sydney. The RAVE web site is at
http://www.rave-survey.org.

\begin{figure}
\centering
\includegraphics[width=12cm]{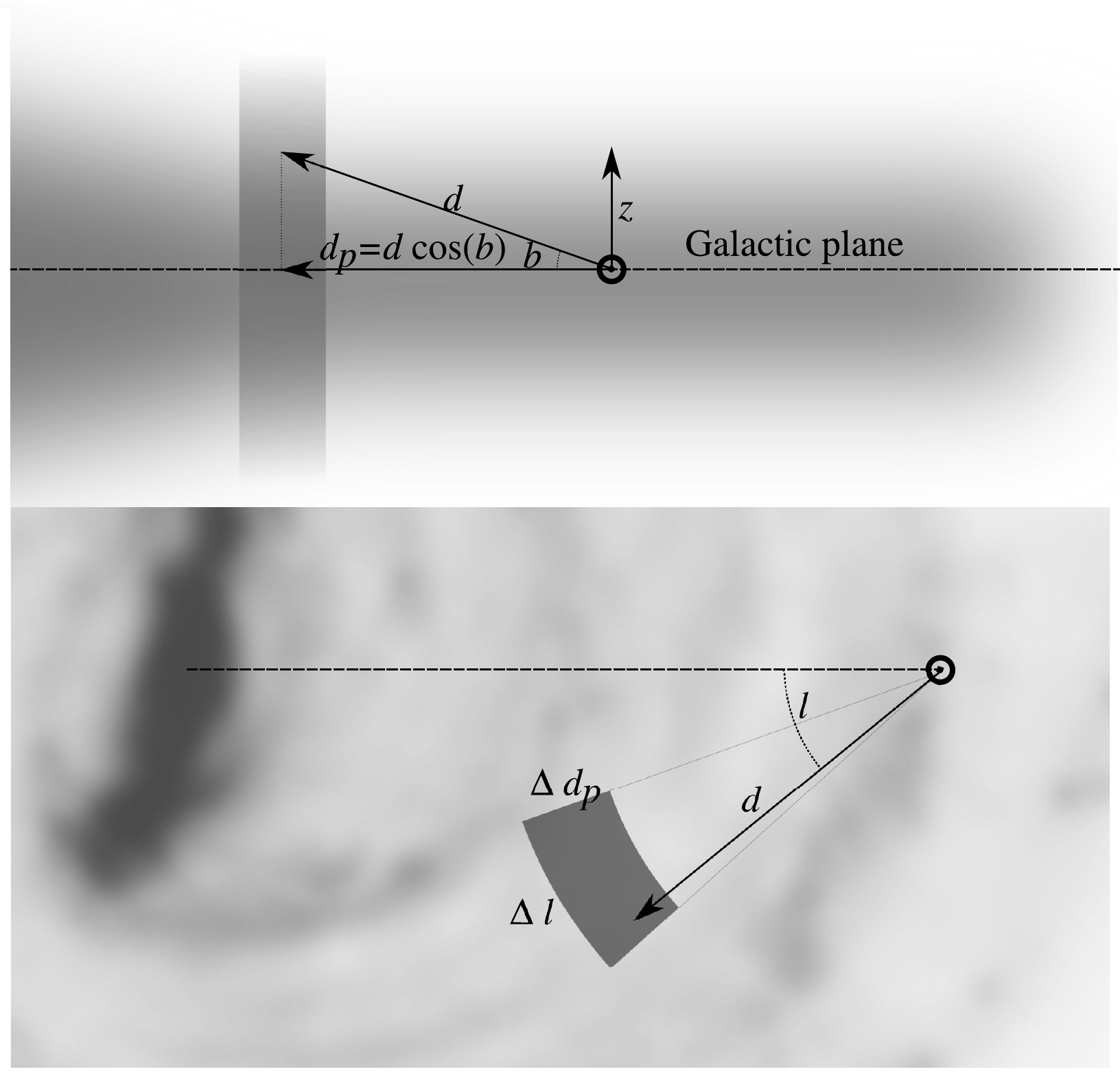}
\caption{\textbf{Coordinate system for spatial sampling of the DIB carrier and dust in the Galaxy.} Top: Side-view of two columns, representing two bins (one stretching from the Galactic plane in the positive $z$ direction and one in the negative $z$ direction), giving two data-points, one for each Galactic hemisphere. The height of the column above (or below) the Galactic plane is limited to 1.5~kpc in $z$ and 40$^\circ$ in  Galactic latitude, primarily to avoid using any spectra that were normalized by a large factor, as the normalization enhances the noise. Because most of the gas and dust is located near the Galactic plane, this limitation should not influence the final results. Bottom: Face-on view of one column, representing one bin, giving one data point. The columns are segments of equally-spaced cylindrical shells. In the first shell, there are three columns, and in each following shell there are $3(2n-1)$ columns, where $n$ is the number of the shell. All the columns thus have the same area of the cross-section and are as close to square as possible. To calculate the maps with different spatial resolutions, only the width of the shell is changed.}
\end{figure}

\begin{figure}
\centering
\includegraphics[width=12cm]{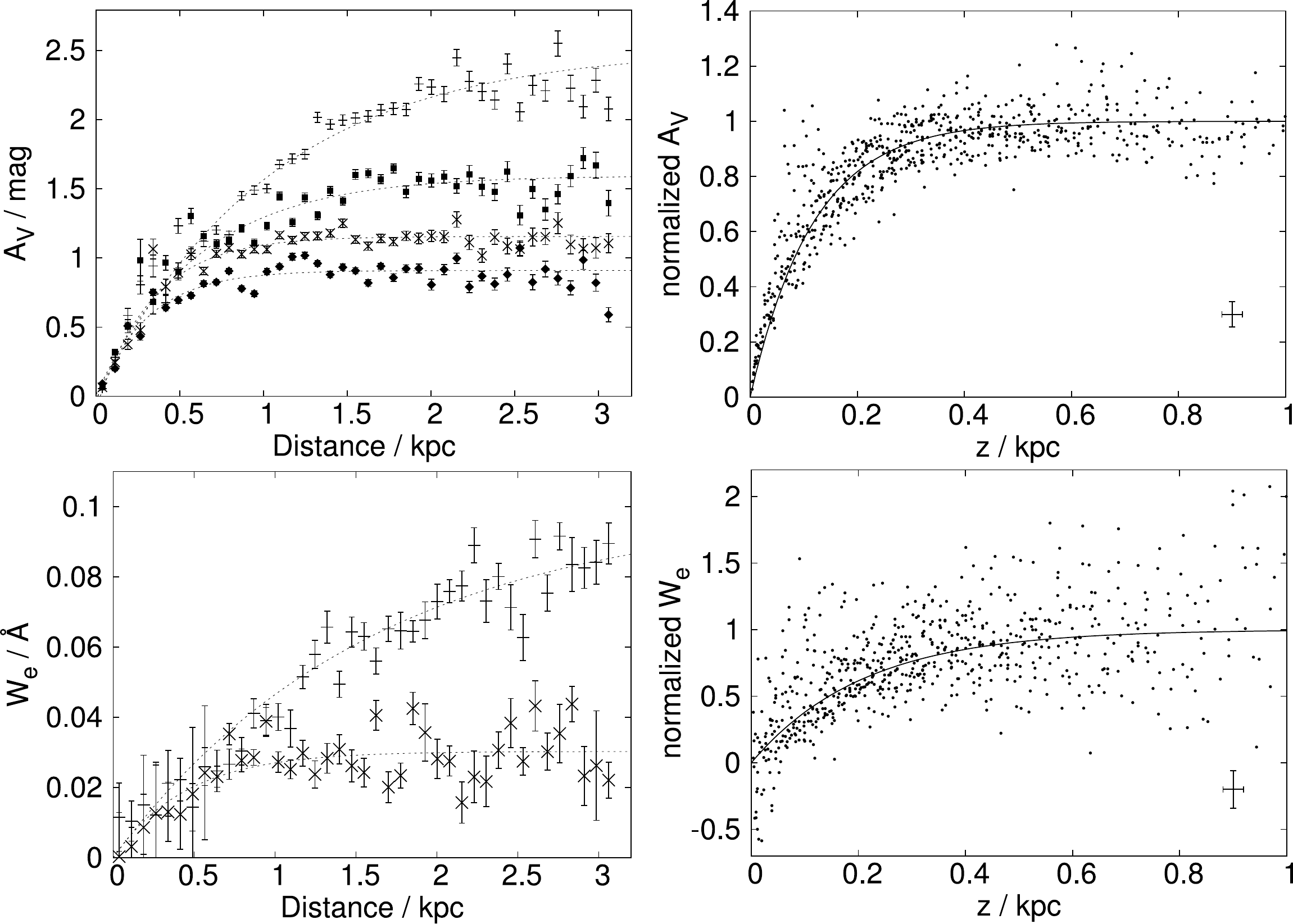}
\caption{\textbf{Determination of the scale-height perpendicular to the Galactic plane.} The average extinction ($A_V$) was calculated by a Bayesian algorithm (top left, \cite{sm}) and equivalent width ($W_e$) of the DIB (bottom left) of stars in 2$^\circ$ wide regions at different central Galactic latitudes ($b$) are shown as a function of distance. The averages extend over all longitudes in every region. Only four examples for the extinction (top to bottom: $b=-6^\circ$, $b=-10^\circ$, $b=-14^\circ$, $b=-18^\circ$) and two for the DIB (top to bottom: $b=-6^\circ$, $b=-14^\circ$) out of 20 regions are plotted here. Dashed curves are fitted exponential models. Data points from all 20 regions are represented in the same plot for the extinction (top right) and for the DIB (bottom right). Instead of the distances $d$, the distance from the Galactic plane 
$z = d \sin(|b|)$ was used, and each was normalized to unity to make the data at all Galactic latitudes comparable. The solid line is the fitted exponential model. Error bars in the bottom right corner represent a typical error of the datapoints.}
\end{figure}

\begin{figure}
\centering
\includegraphics[width=12cm]{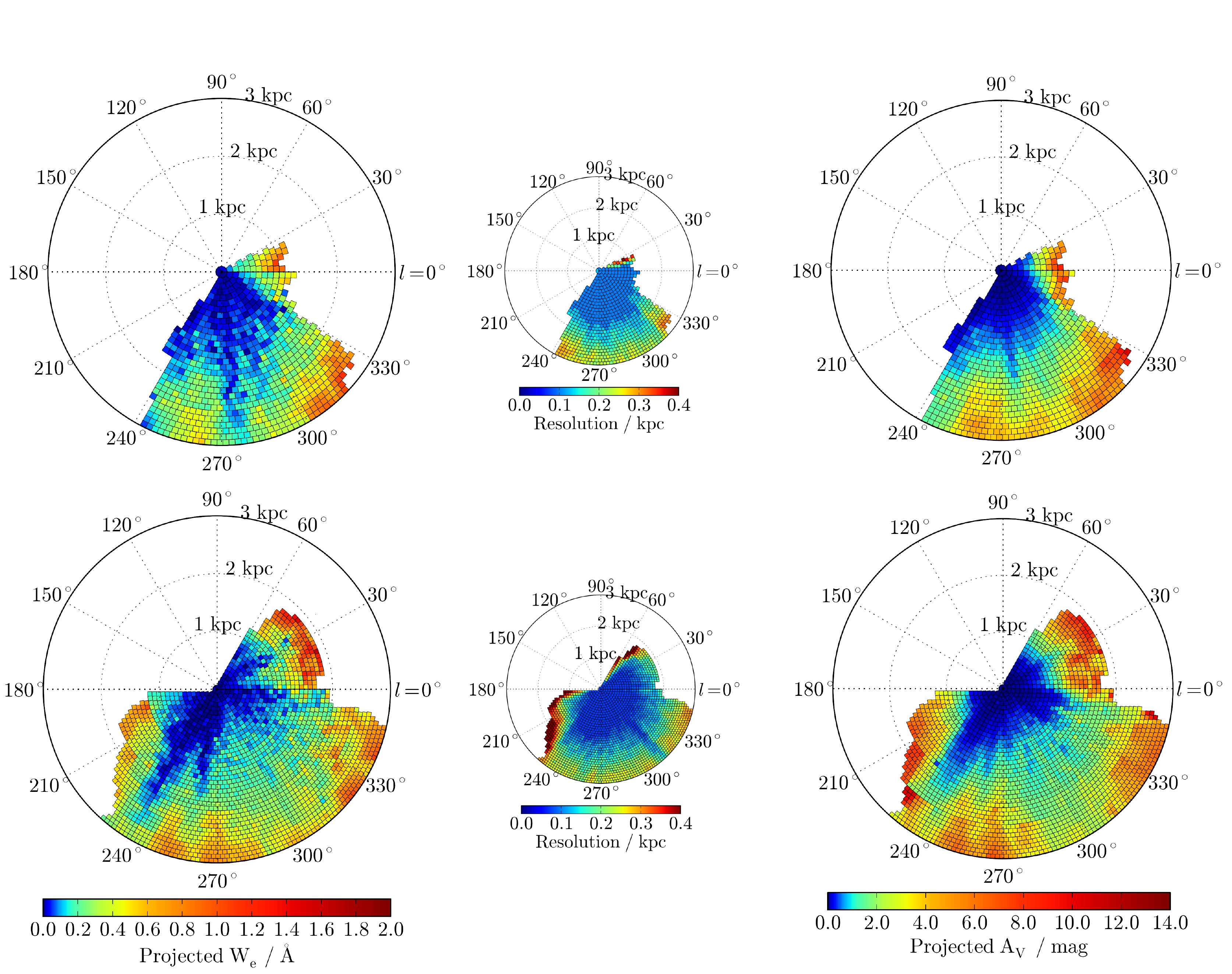}
\caption{\textbf{Projected equivalent width of the DIB at 862~nm, projected extinction and the corresponding spatial resolution of the map.} The large maps show a color-coded projected equivalent width (left) and projected extinction (right). The small maps show the spatial resolution. Top and bottom three maps correspond to the northern and to the southern Galactic hemisphere, respectively. Notice the non-linear color scale. Individual maps with spatial resolution between 0.8 and 0.075~kpc were used to make this combined result \cite{sm}. The Sun is in the center of each map, with projected polar coordinates of the Galactic plane distance and the Galactic longitude. The stars analysed in this study are located south of the celestial equator, which is why Galactic longitude and distance are incompletely sampled between the $l\sim0^\circ$ and $l\sim210^\circ$. There are 1292 bins on the maps for the northern and 2212 bins on the maps for the southern Galactic hemisphere. The typical relative error of the extinction value in each bin is 14\% and the relative error for the $\mathrm{W_e}$ is 12\%.}
\end{figure}

\end{document}